\documentclass[conference]{IEEEtran}
\usepackage{color}
\usepackage{mathpazo}
\usepackage{times}
\usepackage{color}
\usepackage{amsmath}
\usepackage{amsfonts}
\usepackage{latexsym}
\usepackage{amssymb,amsbsy}
\usepackage{cite,url}
\usepackage{mathtools}
\usepackage{amscd}

\usepackage{xcolor,stackengine}

\usepackage{upref}
\usepackage{theorem}
\usepackage{graphicx}
\usepackage{psfrag}
\usepackage{multirow}

\usepackage{mathtools}
\usepackage{xparse}
\DeclarePairedDelimiterX{\set}[1]{\{}{\}}{\setargs{#1}}
\NewDocumentCommand{\setargs}{>{\SplitArgument{1}{;}}m}
{\setargsaux#1}
\NewDocumentCommand{\setargsaux}{mm}
{\IfNoValueTF{#2}{#1} {#1\,\delimsize|\,\mathopen{}#2}}

\DeclarePairedDelimiter\abs{\lvert}{\rvert}
\DeclarePairedDelimiter\ceil{\lceil}{\rceil}

\DeclarePairedDelimiter\parenv{\lparen}{\rparen}

\usepackage{tikz}
\usetikzlibrary{positioning}
\definecolor {processblue}{cmyk}{0.96,0,0,0}
\theoremstyle{plain}
\theorembodyfont{\normalfont\slshape}

\newtheorem{thm}{Theorem$\!$}
\newenvironment{theorem}
{\begin{thm}\hspace*{-1ex}{\bf.}}{\end{thm}}
\newtheorem{lem}[thm]{Lemma$\!$}
\newenvironment{lemma}{\begin{lem}\hspace*{-1ex}{\bf.}}{\end{lem}}

\newtheorem{prop}[thm]{Proposition$\!$}

\newtheorem{cor}[thm]{Corollary$\!$}

\newtheorem{defn}[thm]{Definition$\!$}
\newenvironment{definition}{\begin{defn}\hspace*{-1ex}{\bf.}}{\end{defn}}

\newtheorem{xmpl}[thm]{Example$\!$}
\newenvironment{example}{\begin{xmpl}\hspace*{-1ex}{\bf.}}{\hfill$\Box$\end{xmpl}}

\newtheorem{cnstr}{Construction$\!$}

\renewcommand{\le}{\leqslant}
\renewcommand{\leq}{\leqslant}
\renewcommand{\ge}{\geqslant}
\renewcommand{\geq}{\geqslant}

\newcommand{\N}{\mathbb{N}}
\newcommand{\Z}{\mathbb{Z}}

\newcommand{\eqdef}{\triangleq}
\newcommand{\ccap}{\mathsf{cap}}

\newcommand{\cD}{\mathcal{D}}
\newcommand{\cE}{\mathcal{E}}
\newcommand{\cQ}{\mathcal{Q}}

\begin{document}
\title{Coding for Optimized Writing Rate in DNA Storage}
\IEEEoverridecommandlockouts

\author{
  \IEEEauthorblockN{Siddharth Jain\IEEEauthorrefmark{1}, Farzad Farnoud\IEEEauthorrefmark{2}, Moshe Schwartz\IEEEauthorrefmark{3}, and Jehoshua Bruck\IEEEauthorrefmark{1}}
\IEEEauthorblockA{\IEEEauthorrefmark{1}Electrical Engineering,
California Institute of Technology, U.S.A.,
\texttt{\{sidjain,bruck\}@caltech.edu}}
\IEEEauthorblockA{\IEEEauthorrefmark{2}Electrical \& Computer Engineering,
University of Virginia, U.S.A.,
\texttt{farzad@virginia.edu}}
\IEEEauthorblockA{\IEEEauthorrefmark{3}Electrical and Computer Engineering,
Ben-Gurion University of the Negev, Israel,
\texttt{schwartz@ee.bgu.ac.il}\vspace{-2em}}
\thanks{This work was supported in part by a United States–Israel Binational Science Foundation (BSF) under grant no. 2017652, and by NSF grants under grant nos. CCF-1816409, CCF-1755773, CCF-1816965, and CCF-1717884.}
}

\maketitle

\begin{abstract}
A method for encoding information in DNA sequences is described. The method is based on the precision-resolution framework, and is aimed to work in conjunction with a recently suggested terminator-free template independent DNA synthesis method. The suggested method optimizes the amount of information bits per synthesis time unit, namely, the writing rate. Additionally, the encoding scheme studied here takes into account the existence of multiple copies of the DNA sequence, which are independently distorted. Finally, quantizers for various run-length distributions are designed.
\end{abstract}

\section{Introduction}
DNA is believed to be a compelling medium for storing information due to its high density, energy efficiency, stability and longevity. Studies in the past decade have strengthened this belief by providing evidence that data on DNA can be written, stored and read both outside and inside living organisms~\cite{ChuGaoKos12,GolBerCheDesLepSipBri13,Grass15,ShiNivMacChu17,yazdi2015rewritable,yazdi2017portable,JaiFarSchBru17a}. DNA has been synthesized by a powerful method of phosphoramidite chemistry. However, the quantity and quality of the synthesized DNA using this method is limited by cost~\cite{ChuGaoKos12,GolBerCheDesLepSipBri13} and chemistry limitations~\cite{chm_1,chm_2,chm_3}. 

Recently, a new inexpensive enzymatic method of DNA synthesis was proposed in~\cite{LeeKalGoeBolChu19}. Unlike other synthesis methods~\cite{Synthesis_1,Synthesis_Review} that focus on the synthesis of a precise DNA sequence, in this method information storage does not rely on precision at a single base level. The method relies on the synthesis of runs of homopolymeric bases, the length of which may vary.  Moreover, the length of each run is dependent on the nucleotide appearing before the current run, which makes the run length distribution context dependent or in other words have ``memory". The distribution of run lengths also depends on the synthesis time for each run. One can expect reading long strands as an outcome of this synthesis method which is not a limitation now due to the advance long-read sequencing technologies like Oxford nanopore and PacBio~\cite{longread_1,longread_2}.

In \cite{LeeKalGoeBolChu19}, a coding approach was proposed that encodes information in the transitions between non-identical nucleotides. Hence, any message could be encoded using $3$ symbols (trits as mentioned in the paper). Further, the synthesis method was found to incur deletion, insertion and substitution of nucleotides out of which the deletion error was the most prominent. These errors were tackled by synthesizing multiple strands with the introduction of synchronization bases interspersed between information carrying nucleotides and then using multiple sequence alignment for the reconstruction.

The main drawback in the suggested coding approach in \cite{LeeKalGoeBolChu19} is that it does not make use of the information in the run-length distribution. Thus, the rate of information in \cite{LeeKalGoeBolChu19} is inherently upper bounded by $\log_2{3}$ bits per run.

In this paper, we suggest a method for encoding information that works in tandem with this new synthesis method. As a figure of merit, we study the number of information bits encoded per synthesis time unit, namely, we optimize for the \emph{writing rate}. This differs from previous methods, that optimize the number of bits per channel use. The method we suggest is based on the Precision-Resolution (PR) scheme that was introduced in~\cite{SchBru10} to improve Run-Length Limitation (RLL) constraints. The PR framework provided an optimal set of run lengths when the clock frequency on the transmitter and receiver side was not the same. For the DNA synthesis method described in \cite{LeeKalGoeBolChu19}, the PR framework cannot be applied in its current form~\cite{SchBru10} because of the memory constraint, quantization error of run lengths and the availability of multiple copies of the same message on the receiver end.  We extend the PR framework to include those constraints and design encoder and decoder utilizing the variations in run-lengths to store information in DNA. 


The rest of the paper is organized as follows. In section \ref{sec:pre} we describe constrained systems~\cite{MarRotSie98} and the terminator-free template independent DNA synthesis method introduced in \cite{LeeKalGoeBolChu19}. In section \ref{sec:PR_scheme}, we describe the generalized Precision Resolution scheme and provide examples of encoder and decoder when the run length distributions are Binomial and Poisson. In section \ref{sec:conc}, we conclude the paper and provide directions for future work.
\section{Preliminaries}
\label{sec:pre}

\subsection{Constrained Systems}

Throughout the paper, let $\Sigma$ denote some fixed finite alphabet. We shall eventually take $\Sigma\eqdef\set{A,C,G,T}$ to stand for the four bases found in DNA sequences. We use $\Sigma^*$ to denote the set of all finite strings over $\Sigma$, and $\Sigma^+\eqdef \Sigma^*\setminus\set{\epsilon}$, where $\epsilon$ denotes the unique empty string. We denote the length of $w\in\Sigma^*$ by $\abs{w}$. The concatenation of $w,w'\in\Sigma^*$ is denoted by $ww'$, and $w^i$ denotes a concatenation of $i$ copies of $w$.

Assume $G=(V,E,L)$ is a finite directed and labeled graph, that is $V$ is a finite set of vertices, $E\subseteq V\times V$ is a finite \emph{multiset} of edges (since we allow parallel edges), and $L:E\to \Sigma^+$ are the labels on the edges. We say the \emph{length} of an edge $e\in E$ is $\ell(e)\eqdef\abs{L(e)}$. If all the edge lengths are $1$, we say the graph is \emph{ordinary}.

Let $\gamma=e_1 e_2 \dots e_m$ be a path in $G=(V,E,L)$, $e_i\in E$. We say the path generates the word $L(\gamma)\eqdef L(e_1) L(e_2)\dots L(e_m)$. The empty path generates the empty word $\epsilon$. The graph $G$ is called \emph{lossless} if given vertices $v_1,v_2\in V$ (perhaps the same vertex) and a string $w\in\Sigma^*$, there is at most one path $\gamma$ from $v_1$ to $v_2$ such that $L(\gamma)=w$.

A \emph{constrained system} is represented by a finite directed and labeled graph $G=(V,E,L)$ as above. The language of the constrained system is defined as all the words generated by finite paths in $G$, namely,
\[ S(G)\eqdef \set*{ L(\gamma) ; \text{$\gamma$ is a finite path in $G$}}.\]

The \emph{capacity} of a constrained system $S(G)$ is defined as
\[ \ccap(S(G))\eqdef \limsup_{n\to\infty} \frac{1}{n}\log_2\abs*{S(G)\cap \Sigma^n}.\]
This intuitively measures the average numbers of bits of information per string symbol in long strings. It is well known that there exist encoders with rates below the capacity, and no encoders with rate higher than the capacity. If $G$ is ordinary and lossless, then it is also known that
\[ \ccap(S(G)) = \log_2\lambda(A_G),\]
where $A_G$ is the adjacency matrix of $G$, and $\lambda(A_G)$ is the spectral radius of $A_G$, which by Perron-Frobenius theory is the largest positive eigenvalue of $A_G$. We note that if $G$ is not ordinary, we may easily construct an equivalent ordinary graph $G'$ by converting each edge $v\xrightarrow{w_1\dots w_\ell}u$ (where $w_i\in\Sigma$) into a path $v\xrightarrow{w_1} v_1 \xrightarrow{w_2}\dots \xrightarrow{w_{\ell-1}} v_{\ell-1} \xrightarrow{w_\ell} u$ by inserting auxiliary vertices $v_1,\dots,v_{\ell-1}$.

\subsection{Terminator-Free DNA Synthesis}

In \cite{LeeKalGoeBolChu19}, a method for DNA synthesis is described which is faster and cheaper than known methods. It does, however, suffer from distinct noise characteristics. We describe here the relevant coding-theoretic aspects of this method.

The alphabet is, of course, $\Sigma\eqdef\set{A,C,G,T}$. The synthesis process proceeds in rounds. Assume at the beginning of the round, the current string is $w\in\Sigma^*$. A letter $a\in\Sigma$ is chosen, which differs from the last letter of $w$, denoted $b\in\Sigma$. A chemical reaction is then allowed to occur for a duration of $t\in\N$ time units. The resulting string at the end of the round is $wa^m$, where $m$ is a random variable with distribution $D_{b\to a}(t)$. We emphasize the fact that the distribution of the new run lengths depends on the new letter being appended, the last letter of the string at the beginning of the round, and the duration of the chemical reaction.

\section{Generalized Precision-Resolution Schemes}
\label{sec:PR_scheme}
In this section we generalize the precision-resolution (PR) framework to be able to handle the DNA synthesis method described in Section~\ref{sec:pre}.

The PR framework was introduced in~\cite{SchBru10} as a way to improve RLL constraints. In essence, the RLL constraint was originally devised to solve the problem of clocking differences between the transmitter and receiver. A binary sequence may be thought of as runs of zeros separated by ones, where the information is encoded in the lengths of the runs of zeros. However, if the clock frequency in the transmitter and receiver is not exactly the same, a transmitted run length may be measured differently by the receiver than originally intended by the transmitter. Thus, the constrained code employed should make sure run lengths may be unambiguously recoverable at the receiver side. It was shown in~\cite{SchBru10} that the choice of allowable run lengths in the RLL constraint is sub-optimal, and that the PR framework provides an optimal set of run-lengths, thus increasing the system's capacity.

The PR framework, however, cannot be applied ``as-is'' in the case at hand. This is due to three main differences:
\begin{enumerate}
    \item 
    \textbf{Memory} -- The encoders described in~\cite{SchBru10} are \emph{block encoders}, namely, encoders with a single state only. In the case of DNA synthesis as described in Section~\ref{sec:pre}, we must have memory since the behavior of each writing phase depends on the previous run that was written.
    \item
    \textbf{Quantization error} -- In~\cite{SchBru10} it is assumed that a transmitted run length may change, and that the distribution of the noisy run length has a finite support. Thus, \cite{SchBru10} is able to partition the set of all possible run lengths into disjoint intervals, and the receiver may quantize a received run length to an interval unambiguously. This does not seem to be the case described in~\cite{LeeKalGoeBolChu19}, and we must assume the quantizer at the receiver may err.
    \item
    \textbf{Multiple copies} -- The setting described in~\cite{SchBru10} allows the encoder to create a single output that is transmitted over the channel, and decoded at the receiver. Here, however, multiple copies of the DNA string may be created simultaneously. All of these copies are created from the same user information, but the synthesis process may result in  different DNA strings. We may use these extra copies to improve our overall scheme.
\end{enumerate}
We therefore describe a new framework that builds upon the work of~\cite{SchBru10} and generalizes it.

\subsection{Framework Description}

Fix the alphabet $\Sigma\eqdef\set{A,C,G,T}$. Assume a synthesis round of $t\in\N$ time units, attempting to write a letter $a\in\Sigma$ following a run of $b\in\Sigma$, $b\neq a$. The actual resulting run length is a random variable $r\sim D_{b\to a}(t)$, where we note that the distribution depends on the current run being written, $a$, the previous written run, $b$, and the duration of the synthesis round $t$.

Next, we observe that the synthesis method lends itself naturally to writing several copies of the same string (see~\cite{LeeKalGoeBolChu19}). Let $N\in\N$ denote the number of copies we write. Thus, we have $N$ i.i.d.~random variables $r_1,\dots,r_N$ representing the actual run length resulting in each of the $N$ copies.

Following the general PR framework of~\cite{SchBru10}, we now do the following:
\begin{itemize}
    \item
    We fix a maximal run-length decoding error, $\delta>0$.
    \item
    We fix a maximal synthesis time for all runs, $M$. This is a desirable feature since long runs may affect the DNA molecule's stability.
    \item
    For each $a,b\in\Sigma$, $a\neq b$, a sequence of allowable synthesis-round times,
    $1\leq t_{b\to a}^{(1)} < t_{b\to a}^{(2)} <\dots < t_{b\to a}^{(\ell)}\leq M$,
    to be used when writing a run of $a$'s following a run of $b$'s. Thus, $\ell$ is the number of possible synthesis times for a run . (We note that for later convenience, $\ell$ does not depend on $a$ and $b$.)
    \item    
    Finally, we fix a quantizing function, $\cQ_{b\to a}:\N^N\to [\ell]$. This function receives $(r_1,\dots,r_N)$, a vector of $N$ i.i.d. random variables that are distributed according to $D_{b\to a}(t_{b\to a}^{(i)})$ for some $i$. The function attempts to find $i$, returning $\hat i\in [\ell]$. We require
    {\color{black}$\Pr(\hat i = i|i)\geq 1-\delta$.}
\end{itemize}

{\color{black}We point out here, in the setup described above, we only consider the run-length errors and ignore other types of noise, such as insertion, deletion, and substitution errors. We also observe that this setting may cause a designed run to not appear at all, and then perhaps causing its two adjacent runs to be merged. However, by synthesizing several copies, the probability that the same run is deleted from \emph{all} copies, is exponentially small in the number of copies. A similar strategy was used by a previous work, and run deletions may then be handled by an alignment algorithm~\cite{LeeKalGoeBolChu19}.}

We now construct the following graph $G=(V,E,L)$. We set $V\eqdef \Sigma$. For all $a,b\in\Sigma$, $a\neq b$, we add $\ell$ parallel edges directed from $b$ to $a$ with labels $a^{t_{b\to a}^{(i)}}$, for $i\in[\ell]$.

At this point we make an important observation: the system $S(G)$ does not describe the DNA strings we write. Instead, the strings in $S(G)$ may be thought of as a description of the synthesis algorithm. An edge with a label $a^t$ describes running a synthesis round for the letter $a$ for a duration of $t$ time units. This is in particular important since the capacity, $\ccap(S(G))$ measures the average bits per time unit of synthesis rounds. While this may seem odd at first, we emphasize the following two points. First, the main cost in the transmission process described in this paper is \emph{not} the length of the DNA strings. Instead, the main cost stems from the time it takes to synthesize the DNA molecules. Second, even in a standard use of constrained systems, the length of the transmitted string differs from the length of the received string (for example, due to a difference in clock frequencies). Here, we may think of the transmitter as attempting to write the string describing the synthesis rounds, whereas the receiver reads the resulting DNA string.

\begin{lemma}\label{lem:max_cap}
The graph $G$ constructed in this section is lossless. If $G'$ is the equivalent ordinary graph of $G$ (created by inserting dummy vertices) then
\[\ccap(S(G))=\log_2 \lambda(A_{G'}).\]
\end{lemma}
\begin{IEEEproof}
Assume to the contrary there exist two distinct paths, $\gamma$ and $\gamma'$, from $u\in V$ to $v\in V$ that generate the same word. Without loss of generality we also assume the two paths disagree on the first edge taken.

Let $a^\ell$ be the label of the first edge in $\gamma$. We must have that the first edge of $\gamma'$ is labeled by $a^{\ell'}$, where $\ell\neq\ell'$. Again, without loss of generality, assume $\ell>\ell'$. We must have that $\gamma'$ has a second edge, but by construction, it is labeled by $b^m$ with $b\neq a$. It follows that $L(\gamma)\neq L(\gamma')$, a contradiction. Hence, $G$ is lossless.

The proof for $G'$ follows along the same lines after noting that two hypothetical distinct paths that generate the same word and differ in their first edge, must start in a vertex originally from $G$ (and not in an auxiliary vertex).

Finally, the fact that \[\ccap(S(G))=\ccap(S(G'))=\log_2\lambda(A_{G'}),\]
is well known (e.g., see \cite{LinMar85}).
\end{IEEEproof}

Using standard techniques, e.g., the state-splitting algorithm (see~\cite{LinMar85,MarRotSie98}), we can build encoders $\cE$ for $S(G)$ with rate that asymptotically approaches $\ccap(S(G))$. At the receiver side, the run lengths are qunatized using the quantizing functions $\cQ_{b\to a}$, and a decoder, $\cD$, is employed to reverse the encoding process of $\cE$.

\subsection{Fixing Quantizer Errors}

We now turn to discuss errors due to the quantizers. Intuitively, at the receiver side, the receiver is well aware of the fact that the previous run was a run of $b$'s, whereas the current one is a run of $a$'s, for some $a,b\in\Sigma$, $a\neq b$. It is, therefore, the quantizer's goal to determine $t_{b\to a}^{(i)}$ from the received run length of $a$'s. By construction, the probability of error is at most $\delta$.

Assume a total of $s$ runs was written, where the $j$th run was of designed length $t^{(i_j)}_{a_{i-1}\to a_i}$. It is now our goal to protect the sequence $i_1,i_2,\dots,i_s \in [\ell]^s$ from a typical $\delta$-fraction of errors. To that end, we may use any error-correcting code capable of correcting a typical $\delta$-fraction of errors. Such codes exist with asymptotic redundancy of no more than $s'\eqdef s\parenv{\frac{1}{C_{\delta,\ell}}-1}${\color{black}~\cite{Cover}}, where
\[ C_{\delta,\ell}\eqdef 1+\delta\log_\ell \frac{\delta}{\ell-1}+(1-\delta)\log_\ell(1-\delta). \]
Using such a linear systematic code, we map
$(i_1,i_2,\dots,i_s)\mapsto (i_1,\dots,i_s, i_{s+1},\dots,i_{s+s'})$,
where $i_j\in[\ell]$ for all $j$.

Since $i_1,\dots,i_s$ are already represented by the sequence synthesized thus far, we only need to append DNA bases to represent the redundancy of the code, namely the sequence $(i_{s+1},\dots,i_{s+s'})\in[\ell]^{s'}$. We do so by first identifying our original alphabet $\Sigma$ with $\Z_q$, where $q\eqdef\abs*{\Sigma}$ (where in the case of DNA bases, we get $q=4$). We then transform the sequence of redundancy symbols $(i_{s+1},\dots,i_{s+s'})\in[\ell]^{s'}$ to a sequence over the alphabet $[q-1]=\set{1,2,\dots,q-1}$,
$(i_{s+1},\dots,i_{s+s'})\mapsto(\bar i_1,\dots,\bar i_{s''})$,
where $\bar i_j\in[q-1]$ for all $j$, and
$s''\eqdef\ceil*{s'\cdot \log_{q-1}(\ell)}$.

Finally, to write the redundancy we use runs of designed length $1$, and a $\Z_q$-differential encoding. In detail, if the first $s$ runs were of the letters $a_1,\dots,a_s$ respectively, the following $s''$ runs are of the letters
\[ a_{s+1}=a_s+\bar i_1, a_{s+2}=a_{s+1}+\bar i_2, \dots, a_{s+s''}=a_{s+s''-1}+\bar i_{s''}\]
where addition is done in $\Z_q$.

Overall, upon receiving a transmission, the receiver extracts the sequence of run lengths, $\hat i_1, \dots,\hat i_s$. For the redundancy runs, the run lengths are of no importance, and the letters of the runs determine $\bar i_1,\dots,\bar i_{s''}$, which may be used to find $i_{s+1},\dots,i_{s+s'}$.  The decoder then uses the $\ell$-ary error-correcting code to obtain $i_1,\dots,i_s$ with high probability. It then feeds the corrected sequence to the decoder $\cD$ to obtain the user information. We therefore have:

\begin{theorem}
\label{th:overhead}
Using the setting described in this section, for all large enough $k$, any $k$ user bits may be encoded into a sequence using at most
\[k\cdot \frac{1}{\ccap(S(G))}\cdot \parenv*{1+\parenv*{\frac{1}{C_{\delta,\ell}}-1}\log_{q-1}(\ell)}\]
synthesis time, and be decoded correctly with high probability.
\end{theorem}
\begin{IEEEproof}[Proof sketch]
Take the $k$ bits and feed them to the encoder $\cE$ resulting in a DNA synthesis process of time $k\cdot\frac{1}{\ccap(S(G))}$ asymptotically. In the worst case, this sequence is composed of synthesis rounds of length $1$ only. Protect these $k\cdot\frac{1}{\ccap(S(G))}$ rounds using an error-correcting code with rate (asymptotically) $C_{\delta,\ell}$, resulting in $k\cdot\frac{1}{\ccap(S(G))}\parenv*{\frac{1}{C_{\delta,\ell}}-1}\log_{q-1}(\ell)$ more rounds of length $1$ each that need to be written.
\end{IEEEproof}

Finally, this last theorem may be further improved by assuming the user input is uniform i.i.d.~bits and replacing the worst-case assumption of all rounds of length $1$, with the expected number of rounds. This may be done by a Markov-chain analysis of the encoder $\cE$.

\begin{theorem}\label{thm:main}
Assume the setting described in this section, and let $G'$ be the ordinary version of $G$. Further assume the $k$ user information bits are i.i.d.~uniform random bits. Then for all large enough $k$, the user information bits may be encoded into a sequence using at most
\[k\cdot \frac{1}{\ccap(S(G))}\cdot \parenv*{1+\alpha\parenv*{\frac{1}{C_{\delta,\ell}}-1}\log_{q-1}(\ell)}\]
synthesis time, and be decoded correctly with high probability. Here $\alpha$ is the sum of probabilities of non-auxiliary vertices in the stationary distribution of the max-entropic Markov chain over $G'$.
\end{theorem}
\begin{IEEEproof}[Proof sketch]
The proof is similar to that of Theorem~\ref{th:overhead}, with the difference being that the expected number of synthesis rounds (after encoding the user information bits) is $k\cdot\frac{1}{\ccap(S(G))}\cdot\alpha$. This is due to the fact that a new round starts every time we reach a non-auxiliary vertex in $G'$.
\end{IEEEproof}

\subsection{Quantizers for Binomial Run Lengths}
We proceed in this section and the following one, to study some special cases of quantizers, depending on the distribution of run lengths. In this section we assume that the run lengths have binomial distribution. Further, we assume that $D_{b\to a}(t)$  are independent of the choice of $a$ and $b$, i.e. $D_{b\to a}(t) = D(t)$.

\begin{figure*}[h]
\centering
(a)\scalebox{0.33}{\includegraphics{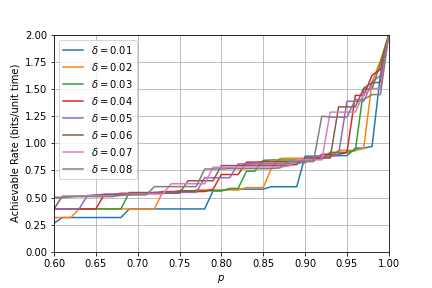}}(b)\scalebox{0.33}{\includegraphics{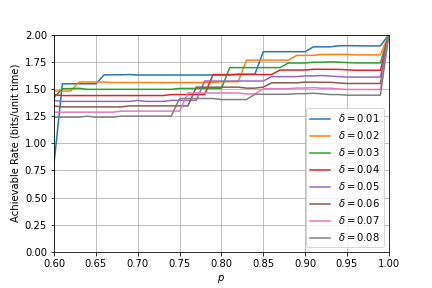}}(c)\scalebox{0.33}{\includegraphics{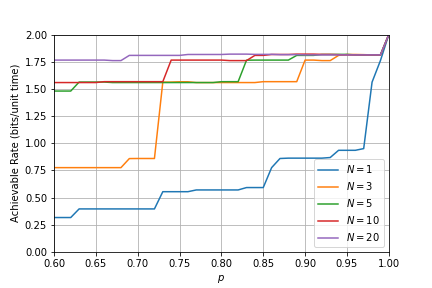}}
\caption{(a) Achievable rate for $\mathrm{Binomial}(Nt,p)$, $N = 1$ and $M= 10$ for different choices of $\delta$ using Theorem \ref{thm:main}. (b) Achievable rate for $\mathrm{Binomial}(Nt,p)$, $N = 5$ and $M = 10$ for different choices of $\delta$ using Theorem \ref{thm:main}. (c) Comparison of Achievable rates for $ \mathrm{Binomial}(Nt,p)$, $\delta = 0.02$ and $M = 10$ for different choices of $N$ using Theorem \ref{thm:main}.}
\label{fig:PR_delta_N}
\end{figure*}

We construct a maximum-likelihood quantizer to find the correct run-length index. For a given run, let $(r_i)_{1\leq i\leq N}$ denote the run lengths received at the output from $N$ copies that were synthesized. Assuming $r_i$ is i.i.d and distributed according to $\mathrm{Binomial}(t,p)$,  the received run length sum $r(t) = 
\Sigma_{i=1}^Nr_i$ is  distributed according to $\mathrm{Binomial}(Nt,p)$, where $t \in \{t^{(1)},t^{(2)},\cdots, t^{(\ell)}\}$, i.e. 
\[\Pr(r(t) = k) = \binom{Nt}{k}p^k{(1-p)}^{Nt-k}.\] 
The log-likelihood is given by $$\log \Pr(r|t^{(j)}) = \log \binom{Nt^{(j)}}{r}+r\log \frac{p}{1-p}+Nt^{(j)}\log(1-p).$$ The index $j^*$ is determined by $$j^* = \arg \max_j \log \binom{Nt^{(j)}}{r}+r\log \frac{p}{1-p}+Nt^{(j)}\log(1-p).$$ Let the true index be denoted by $i$. Let $\tau_h,\ h=0,\dotsc,\ell$, be such that the range for which the decision is correct, i.e., $j^*=i$, is given by $$\tau_{i-1}< r\leq \tau_{i}.$$ The probability of error is  {\color{black}$$\Pr(j^* \neq i|i) = \Pr(r > \tau_i)+\Pr(r \leq \tau_{i-1}) \leq \delta.$$}
Below, we describe the steps to design an encoder for this setup for different values of $p$ and $\delta$. 
\begin{itemize}
    \item {\it Step 1:}   Set $\tau_0=0$ and 
 \begin{equation*}
 \begin{split}
     t^{(1)}&=\min\{t:\Pr(r(t)>0)\ge1-\delta\}\\
     \tau_1 &= \min\{x:\Pr(1\leq r(t^{(1)})\leq x) \geq 1-\delta\}
 \end{split}
 \end{equation*}
 
    \item {\it Step 2: } For each $2\leq i \leq \ell$, 
    $t^{(i)}$ is calculated by finding the minimum $t>t^{(i-1)}$ such that 
    \begin{align*}
    \log\frac{(Nt^{(i-1)}+1)\cdots Nt}{(Nt^{(i-1)}-\tau_{i-1}+1)\cdots (Nt-\tau_{i-1})}\\+(Nt-Nt^{(i-1)})\log(1-p) \leq 0.
    \end{align*}
    and $\tau_{i} = \min\{x:\Pr(\tau_{i-1}< r(t^{(i)})\leq x) \geq 1-\delta\}.$
    
    Note that $\ell$ is the index where $t^{(\ell)}\leq M$ and $t^{(\ell+1)} > M$.
\end{itemize}
Figures \ref{fig:PR_delta_N}(a) and \ref{fig:PR_delta_N}(b) show the achievable rates for the PR system described above for $N = 1$ and $N = 5$, respectively, for different choices of $p$, $\delta$, and $M = 10$. Figure \ref{fig:PR_delta_N}(c) shows the achievable rates for fixed parameters, except for varying values of $N$.
\begin{figure*}[h]
\centering
(a)\scalebox{0.4}{\includegraphics{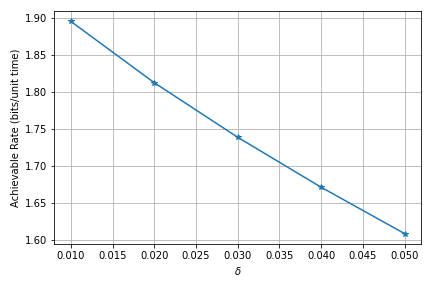}}(b)\scalebox{0.4}{\includegraphics{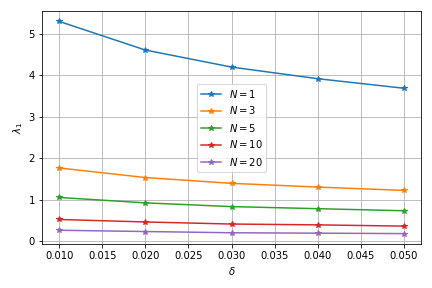}}
\caption{(a) Achievable rates for Poisson Run lengths for different values of $\delta$ using Theorem \ref{thm:main} with $\lambda_j = \lambda_1{(t^{(j)})}^2$ where $t^{(1)} = 1$. Here $\lambda_1$ (see Figure \ref{fig:PR_Poisson}(b)) is a function of $N$ and $\delta$. (b) Required $\lambda_1$ for different values of $N$ and $\delta$.} 
\label{fig:PR_Poisson}
\end{figure*}
\subsection{Quantizers for Poisson Run Lengths}

Suppose that the run lengths have a Poisson distribution with a parameter that can be set by the encoder. The parameter is chosen from the set  $\{\lambda_1,\lambda_2,\dotsc\}$, $\lambda_j<\lambda_{j+1}$. 
The decoder decides $\lambda_{j^*}$ if $\tau_{j^*-1}<\bar r  \leq \tau_{j^*}$ where $\bar r = \frac{1}{N}\sum_{i=1}^Nr_i$,  
and $\{\tau_0,\tau_1,\cdots\}$ are designed such that\[\Pr(\bar r \leq \tau_{j^*-1}|\lambda_{j^*}) \leq \frac{\delta}{2} \quad\text{and}\quad
\Pr(\bar r > \tau_{j^*}|\lambda_{j^*}) \leq \frac{\delta}{2}.\]
Let the true index be denoted by $i$. 
The range for which the decision is correct is given by
\[
\tau_{i-1}<\bar r\leq\tau_i
\]
The probability of error is given by:  
    \[\Pr(\bar r> \tau_i)+\Pr(\bar r\le \tau_{i-1})\le \delta\] 
for $N\bar r\sim Poisson(N\lambda_i).$

Below, we describe the design an encoder for this setup. 
\begin{itemize}
    \item \emph{Step 1:} Set $\tau_0 = 0$ and $\lambda_1 = \frac{1}{N}\log(\frac{2}{\delta}).$
 \item \emph{Step 2:}  For each $1\leq i \leq \ell-1$, $\tau_i$ and 
    $\lambda_{i+1}$ are calculated as follows: 
    \[\tau_i = \min\{x:\Pr(\bar r> x) \leq \frac{\delta}{2}\}\]
 for $N\bar r\sim \mathrm{Poisson}(N\lambda_i)$.
 \[\lambda_{i+1} = \min\{\lambda > \lambda_i:\Pr(\bar r \leq\tau_i) \leq \frac{\delta}{2}\}\]
 for $N\bar r\sim \mathrm{Poisson}(N\lambda)$.
\end{itemize}

Figure \ref{fig:PR_Poisson}(a) shows the achievable rate for different values of $\delta$ for Poisson run length based encoder and decoder obtained using the steps described above with $\ell = 10$. Figure \ref{fig:PR_Poisson}(b) shows the $\lambda_1$ required for different values of $N$ to achieve the rate curve given in Figure \ref{fig:PR_Poisson}(a). 

{\color{black}We point out here that the steps presented for choosing $t^{(1)},t^{(2)},\cdots,t^{(\ell)}$ in sections III-C and III-D are based on heuristics and may not be optimal. Nevertheless, the results show that the achievable rate is well beyond the maximum rate of $\log_2 3$ of~\cite{LeeKalGoeBolChu19}.}
\section{Conclusion}
\label{sec:conc}

We described in this paper a method for encoding information in DNA sequences, that is based on the precision-resolution framework \cite{SchBru10}, in conjunction with the terminator-free synthesis method \cite{LeeKalGoeBolChu19}. This method takes into account several possible run lengths for each run of the synthesized sequence, thus increasing the rate beyond the fundamental upper bound of $\log_2 3$ bits per symbol. Additionally, the method accounts for quantizer error during the reading process, and it utilizes the fact that the receiver is in possession of multiple copies (independently distorted) of the synthesized sequence. {\color{black}We make note that the synthesis process may delete entire runs, perhaps also causing neighboring runs to merge. However, as in a previous work~\cite{LeeKalGoeBolChu19}, such deletions may be handled by the alignment algorithm.}

We also studied Binomial and Poisson distributions for synthesized run lengths, and provided methods for designing quantizers for these distributions. A point for future research is to conduct empirical studies of the synthesis process of \cite{LeeKalGoeBolChu19} in order to find the actual parameters of run length distributions.

Finally, we believe this method should eventually be used in conjunction with addressless information encoding, such as coding of sliced information \cite{SimRavBru19}. We leave this for future research.

\bibliographystyle{IEEEtranS}
\bibliography{allbib}
\end{document}